\documentclass[letter]{aa}

\usepackage{graphicx}
\usepackage{natbib}
\usepackage{txfonts}
\begin{document}

   \title{Distance to three molecular clouds in the central molecular zone}

  \author{F. Nogueras-Lara
          \inst{1}
          \and      
          R. Sch\"odel 
          \inst{2}       
          \and
          N. Neumayer 
          \inst{1} 
          \and
          M. Schultheis 
          \inst{3}                 
          }

   \institute{
    Max-Planck Institute for Astronomy, K\"onigstuhl 17, 69117 Heidelberg, Germany
              \email{nogueras@mpia.de}
       \and 
           Instituto de Astrof\'isica de Andaluc\'ia (IAA-CSIC),
     Glorieta de la Astronom\'ia s/n, 18008 Granada, Spain
       \and 
           Université Côte d'Azur, Observatoire de la Côte d'Azur, Laboratoire Lagrange, CNRS, Blvd de l'Observatoire, F-06304 Nice, France                                    
       }
   \date{}

 
  \abstract
   {The determination of absolute and relative distances of molecular clouds along the line-of-sight towards the central molecular zone (CMZ) is crucial to infer its orbital structure, dynamics, and to understand star formation in the clouds.}
   {Recent results by \citet{Zoccali:2021aa} suggest that the G0.253 + 0.016 cloud (the Brick) does not belong to the CMZ. This motivated us to cross check their results computing the absolute and relative distance to the Brick and also to other two molecular clouds (the 50 km/s, and the 20 km/s clouds), and discuss their CMZ membership.} 
    {We used the colour magnitude diagrams $K_s$ vs. $H-K_s$ to compare stars detected towards the target clouds with stars detected towards three reference regions in the nuclear stellar disc (NSD) and the Galactic bulge. We used red clump (RC) stars to estimate the distance to each region.}
   {We obtained that all the clouds present a double RC feature. Such a double RC has been reported in previous work for the NSD, but not for the bulge adjacent to it. We exclude the possibility that the different RC features are located at significantly different distances. The obtained absolute and relative distances are compatible with the Galactic centre distance ($\sim8$\,kpc).}

   {}


   \maketitle
%

\section{Introduction}

The central molecular zone (CMZ) is an accumulation of dense molecular gas (R$\lesssim$200\,pc), where the gas channeled  through the Galactic bar settles in orbits around the Galactic centre (GC) \citep[e.g.][]{Morris:1996vn}. It contains $\sim3-5\times 10^7$\,M$_\odot$ of molecular gas \citep{Dahmen:1998aa,Pierce-Price:2000aa} and constitutes a star-forming region characterised by extreme conditions such as a high confining pressure of $10^{6 - 7}$\,K\,cm$^{-3}$ \citep{Yamauchi:1990aa,Spergel:1992aa,Muno:2004aa}, and strong magnetic fields \citep[$\sim5$\,mG, ][]{Pillai:2015aa}. The CMZ partially overlaps with the nuclear stellar disc (NSD), a distinct disc-like structure of stars \citep{Launhardt:2002nx} characterised by mainly old stars \citep[$>$80\,\% of the stellar mass formed $>8$\,Gyr ago, ][]{Nogueras-Lara:2019ad}. The presence of young clusters and HII regions highlights recent and ongoing star formation in the NSD, being the Galaxy's most prolific (massive) star forming region when averaged by its volume  \citep[e.g. the Arches and the Quintuplet clusters][]{Figer:1999uq,Figer:2002qf,Schneider:2014vn}. Nevertheless, the currently observed star formation in the CMZ is about an order of magnitude below what is expected based on the amount of dense gas \citep[e.g.][]{Longmore:2013ab,Kauffmann:2017aa}.

The determination of the absolute and the relative distance between the clouds forming part of the CMZ is crucial to infer its structure and dynamics, and to better understand how star formation proceeds in this environment \citep[e.g.][]{Kruijssen:2015aa}. Its 3D structure has not been fully constrained yet and there are several possible scenarios ranging from a twisted elliptical ring \citep{Molinari:2011fk}, over two spiral arms \citep[e.g.][]{Ridley:2017aa}, to an open gas streams \citep{Kruijssen:2015aa}. In this sense, it is of fundamental interest to distinguish between dark clouds belonging to the CMZ and possible foreground clouds \citep{Longmore:2013ab}.

Observations along the line-of-sight towards the GC are hampered by extreme extinction ($A_V\gtrsim30$ mag, $A_{K_s}\gtrsim2.5$ mag, \citealt[e.g.][]{Nishiyama:2008qa,Schodel:2010fk,Nogueras-Lara:2018aa,Nogueras-Lara:2020aa}), and crowding. Given these limitations, the majority of stars that can be observed are red giants. In particular, red clump (RC) stars \citep[metal rich, core helium burning giants, e.g.][]{Girardi:2016fk} are the dominant and best tracer of the structure in the inner Galaxy, and can be used to estimate distances towards the CMZ molecular clouds \citep[e.g.][]{Longmore:2012uq,Zoccali:2021aa}.

In this letter, we estimate the distance to three molecular clouds, G0.254+0016 \citep[the "Brick",][]{Longmore:2012uq} , the 50 km/s cloud and the 20 km/s cloud \citep[e.g.][]{Kauffmann:2017aa}, and discuss whether they belong to the CMZ. In particular, the recent paper by \citet{Zoccali:2021aa} found that the Brick cloud is not located at the GC, so we aimed at cross-validating their results.

\section{Data}

To overcome the observational challenges, we used $H$ and $K_s$ photometry from the GALACTICNUCLEUS (GNS) survey \citep{Nogueras-Lara:2018aa,Nogueras-Lara:2019aa}. The GNS survey offers $JHK_s$ photometry of the inner GC at high angular resolution ($\sim0.2"$). It also suffers little from saturation. Therefore it offers a high dynamic range and is complete below the RC, indispensable features for GC research. Statistical photometric uncertainties are $\lesssim0.04$\,mag at the RC. The systematic uncertainty of the zero points (ZPs) is 0.04\,mag in each band.

\begin{figure}
   \includegraphics[width=\linewidth]{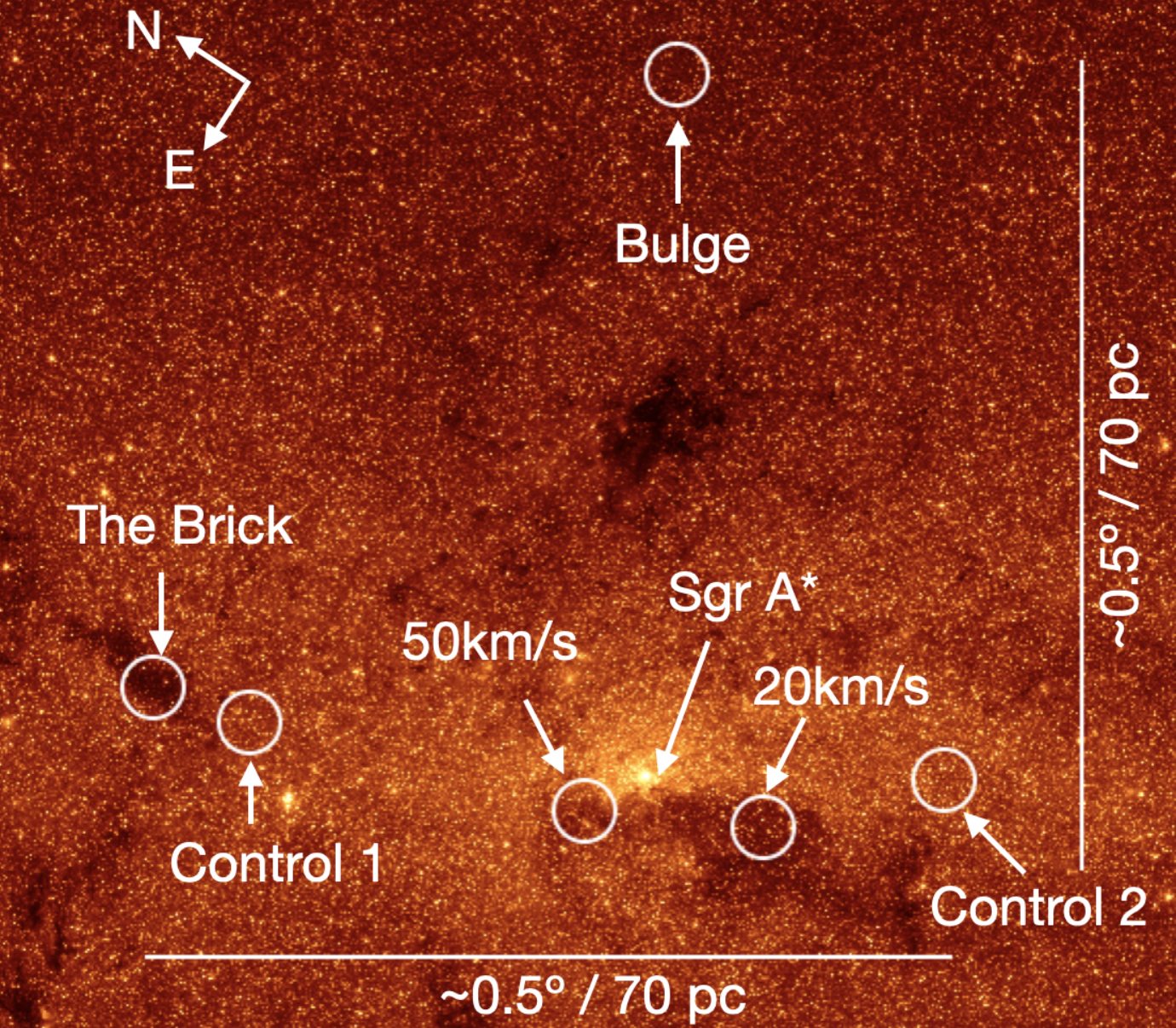}
   \caption{Target regions over plotted on a Spitzer/IRAC image at 3.6\,$\mu$m.}

   \label{GNS}
    \end{figure}

For our analysis, we chose circular regions with a radius of $1.2'$ centred on the coordinates shown in Table \ref{regions}. Three of the regions are centred on the target clouds, the other three on control regions. The control regions are two low-extinction fields in the NSD and one in the nearby bulge. Figure\,\ref{GNS} shows the location of the regions.

\begin{table}
\caption{Target regions.}
\label{regions} 
\begin{center}
\def\arraystretch{1.3}
\setlength{\tabcolsep}{3.8pt}

\begin{tabular}{ccc}
\hline 
\hline 
Region & $l$ & $b$\tabularnewline
 & ($^\circ$ $'$ $''$) & ($^\circ$ $'$ $''$) \tabularnewline
\hline 
50 km/s & +359:58:48.019 & -0:04:12.019\tabularnewline
20 km/s & +359:52:12.022 & -0:04:48.016\tabularnewline
The Brick & +0:14:43.879 & +0:00:23.670\tabularnewline
Control 1 & +0:11:09.816 & -0:00:54.320\tabularnewline
Control 2 & +359:45:30.495 & -0:03:01.103\tabularnewline
Bulge & +359:55:22.762 & +0:22:59.487\tabularnewline
\hline 
\end{tabular}

\end{center}

 \end{table}

The Brick field images were of lower than mean quality in the GNS survey. Therefore the conservative settings of the pipeline were not adequate for this field. For this reason, we performed dedicated PSF fitting photometry using {\it StarFinder} \citep{Diolaiti:2000qo} on the corresponding GNS image.

\begin{figure*}
   \includegraphics[width=\linewidth]{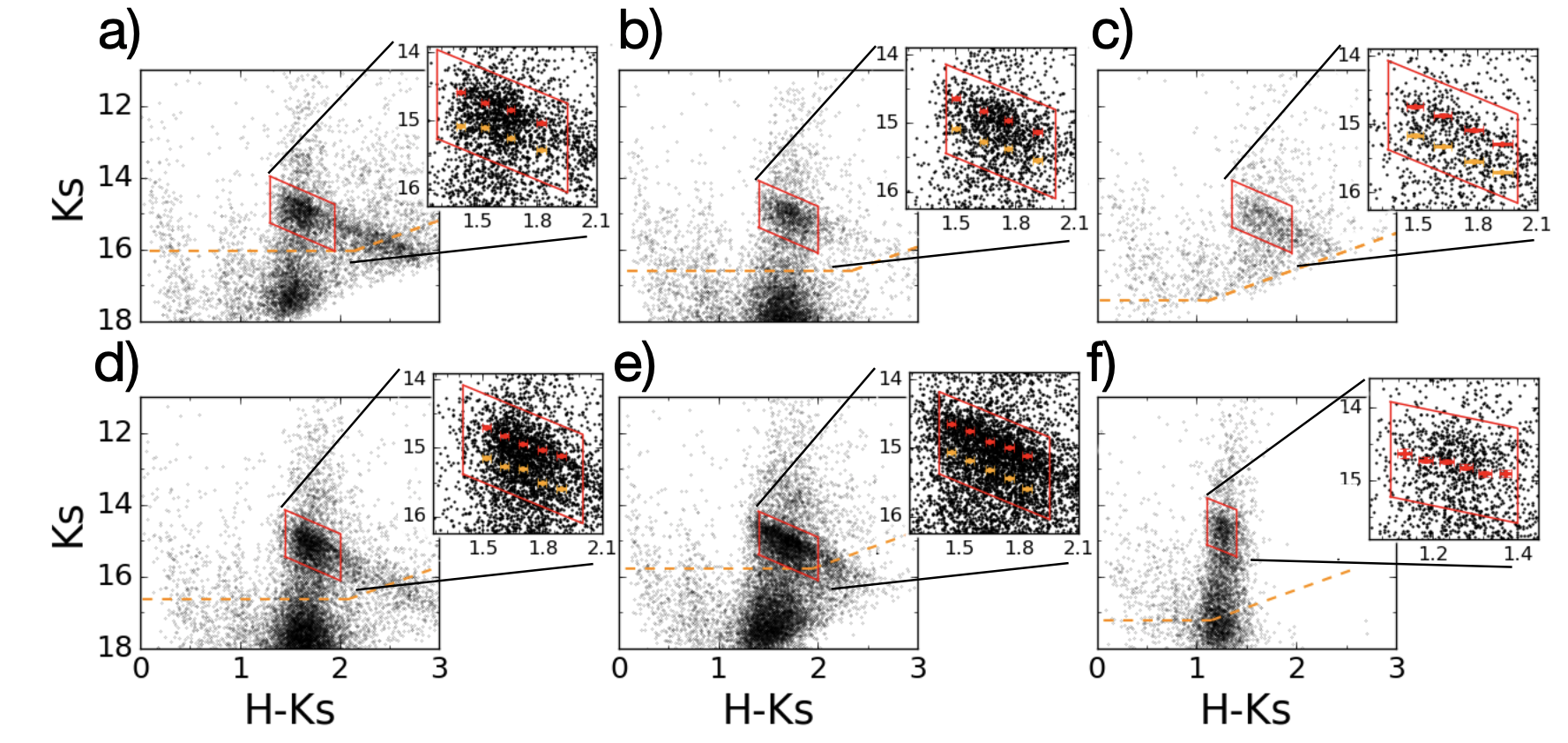}
   \caption{Colour magnitude diagrams $K_s$ vs. $H-K_s$ of the target regions: a) 50 km/s cloud. b) 20 km/s cloud. c) The Brick cloud. d) Control 1. e) Control 2. f) Bulge region. The red parallelograms show the RC feature and the cut applied for the foreground population. The insets are a zoom-in on the RC regions where we detect a double RC (Sect.\,\ref{sequences}). The red and orange points correspond to the best fit obtained for the two RC features and indicate the 1\,$\sigma$ uncertainties. The orange dashed lines mark the 80\,$\%$ completeness level.}

   \label{CMDs}
    \end{figure*}

We estimated the completeness due to crowding (completeness at the magnitudes of interest is not limited by sensitivity in the crowded GC fields), in the same way as it was derived for the GNS survey in Sect.\,4.3. of \citet{Nogueras-Lara:2019aa}. Figure\,\ref{CMDs} shows the colour magnitude diagrams (CMDs) $K_s$ vs. $H-K_s$ of the target regions, where the 80\,\% completeness limit is indicated for each of the regions. 

Given the relatively small size of the target regions, the contamination from foreground population (stars in the line-of-sight from the Earth to the GC mainly from the Galactic disc) is not significant. Nevertheless, we excluded stars with $H-K_s<1.3$\,mag for the regions in the NSD \citep[removing stars belonging to the disc and the bulge,][]{Nogueras-Lara:2019ad,Sormani:2020aa}. In the case of the bulge region, we applied a colour cut $H-K_s\sim1.1$\,mag, in agreement with previous work \citep{Nogueras-Lara:2018ab}.

 \section{Analysis of the RC features}
 
 \subsection{RC features and star formation}
 \label{sequences}
 
The analysis of the RC features in the CMD can give us insights into the star formation history (SFH) of the studied regions. In particular, the NSD is characterised by a double RC feature associated with its dominant old stellar population (bright  RC), and a significant star formation event $\sim1$\,Gyr ago ($\sim5\,\%$ of the total stellar mass), corresponding to the faint RC \citep[see Fig.\,1 in][]{Nogueras-Lara:2019ad}. On the other hand, the innermost regions of the Galactic bulge are characterised by an alpha enhanced, metal rich, old stellar population \citep[e.g.][]{Fulbright:2006aa,Fulbright:2007aa,Babusiaux:2010aa,Hill:2011aa,Bensby:2011aa,Ness:2013aa,Johnson:2013aa,Bensby:2017aa,Rojas-Arriagada:2020aa} that produces a single RC feature \citep{Nogueras-Lara:2018ab}. Thus, studying the RC feature for a given cloud, it is possible to infer whether the majority of stars in front of the cloud belong to the NSD or to the foreground population (mostly inner bulge as in our bulge control field).

We analysed the RC features inside the red parallelograms in Fig.\,\ref{CMDs}. We applied the techniques described in  \citet{Nogueras-Lara:2018ab,Nogueras-Lara:2019ac,Nogueras-Lara:2020aa} to distinguish between an individual or a double RC in the data. Namely, we divided the features into colour bins ($\sim0.1$\,mag depending on the density of the points for each region) and fit a single- and a double-Gaussian model to the $K_s$ distribution for each colour bin. We used the Akaike information criterion (AIC) \citep{Akaike:1974aa} to compare both models. We obtained that a double RC model is favoured in all cases, but in the one corresponding to the inner bulge, consistent with our previous work on the entire inner GC region \citep{Nogueras-Lara:2019ad,Nogueras-Lara:2020aa}. The inner bulge region only shows a single RC in agreement with its mainly old stellar population \citep{Nogueras-Lara:2018ab}. Table\,\ref{table2} shows the distances between features obtained for each case. The statistical uncertainties were estimated using a jackknife resampling method, and the systematics were obtained by varying the bin width, and the outlines of the parallelogram to select the stars.

\begin{table*}
\caption{Analysis of the RC features.}
\label{table2} 
\begin{center}
\def\arraystretch{1.3}
\setlength{\tabcolsep}{3.8pt}

\begin{tabular}{ccccccc}
\hline 
\hline 
Region & $\Delta K_s$ & $s_1$ & $cut_1$ & $s_2$ & $cut_2$ & distance\tabularnewline
 & (mag) &  & (mag) &  & (mag)& (kpc)\tabularnewline
\hline 
50 km/s & 0.42$\pm$0.05$\pm$ 0.02  & 1.10$\pm$0.08$\pm$0.05 & 13.03$\pm$0.08$\pm$0.13 & 0.92$\pm$0.61$\pm$0.11 & 13.72$\pm$0.49$\pm$0.20& 8.4$\pm$0.3$\pm$0.6\tabularnewline
20 km/s & 0.43$\pm$ 0.01$\pm$ 0.02 & 1.18$\pm$0.13$\pm$0.08 & 12.89$\pm$0.16$\pm$0.17 & 1.10$\pm$0.13$\pm$0.13 & 13.44$\pm$0.14$\pm$0.22& 7.9$\pm$0.6$\pm$0.7\tabularnewline
The Brick & 0.45$\pm$ 0.03$\pm$ 0.01 & 1.30$\pm$0.21$\pm$0.06 & 12.78$\pm$0.28$\pm$0.09 & 1.14$\pm$0.34$\pm$0.14 & 13.49$\pm$0.36$\pm$0.23& 7.6$\pm$1.0$\pm$0.4\tabularnewline
Control 1 & 0.46$\pm$ 0.05$\pm$ 0.04 & 1.09$\pm$0.18$\pm$0.06 & 13.07$\pm$0.19$\pm$0.15 & 1.22$\pm$0.08$\pm$0.07 & 13.29$\pm$0.10$\pm$0.14& 8.6$\pm$0.8$\pm$0.7\tabularnewline
Control 2 & 0.44$\pm$ 0.02$\pm$ 0.05 & 1.22$\pm$0.07$\pm$0.08 & 12.86$\pm$0.09$\pm$0.16 & 1.36$\pm$0.07$\pm$0.05 & 13.08$\pm$0.09$\pm$0.11& 7.8$\pm$0.3$\pm$0.7\tabularnewline
Bulge & - & 1.20$\pm$ 0.20$\pm$0.13 & 13.27$\pm$0.24$\pm$0.14 & - & - &9.5$\pm$1.0$\pm$0.7\tabularnewline
\hline 
\end{tabular}

\end{center}

\footnotesize
\textbf{Notes.} $\Delta K_s$ indicates the distance between the two RCs obtained. $s_i$ and $cut_i$ are the slopes and the intercepts of the bright ($i=1$) and faint ($i=2$) features. Distance indicates the distance The uncertainties correspond to the statistical and systematic ones, respectively. 

 \end{table*}

\subsection{Reddening vector and distances}

The colour-cut applied to our data guarantees the exclusion of stars from the Galactic disc. As pointed out by \citet{Zoccali:2021aa}, any RC feature significantly closer to Earth than the GC would manifest itself as an over-density vertically offset from the reddening line fit to the GC RC feature in the CMDs. We estimate that a difference of 1\,kpc at the GC distance, implies a variation in $K_s$ of $\sim0.3$\,mag. Therefore, the analysed RC features do not include any significant contribution from stars at different distances, given that there are not any points significantly out of the general slope trend in the RC features, within the uncertainties. In this way, the slopes of the RC features indicate the direction of the reddening vector and can be translated into the extinction curve \citep[e.g.][]{Nishiyama:2009oj,Nogueras-Lara:2020aa}. We obtained the reddening free magnitude for the RC features computing the corresponding linear fit and following the equation:

\begin{equation}
\label{equation_cut}
K_{s0} = s\times(H-K_s)_0+cut \hspace{0.5cm},
\end{equation}

\noindent where $K_{s0}$ is the reddening free magnitude, $s$ is the slope of the reddening vector, $(H-K_s)_0=0.07\pm0.03$\,mag is the intrinsic colour of the RC \citep{Nogueras-Lara:2019ad}, and $cut$ is the intercept of the linear fit.

We applied a linear fit to the each of the obtained RC features. Table\,\ref{table2} shows the results, where the uncertainties were estimated as previously explained. Moreover, we considered the contribution of the systematic uncertainty of the ZPs to the uncertainties of the estimated intercepts, and added it quadratically to the systematics. The resulting uncertainty of the intercepts is  $\sim$0.1\,mag, and was added quadratically to the systematics. We converted the reddening free magnitudes into distances using the distance modulus and assuming an absolute magnitude of $M_{K_s} =-1.59\pm0.04$ for RC stars, computed averaging over the values obtained by \citet{Groenewegen:2008wj,Hawkins:2017aa,Chan:2020ab}. The uncertainty corresponds to the standard deviation of the measurements. We applied a population correction of $\Delta M_K = -0.07\pm0.07$, as explained in \citet{Schodel:2010fk}. We only used the bright RC to derive the distances, given the larger uncertainties of the faint one and the difficulty of determining a proper $M_{K_s}$ given that the RC brightness is very sensitive to age variations for stellar populations $\sim1$\,Gyr \citep[see Fig.\,6 in][]{Girardi:2016fk}. The last column of  Table\,\ref{table2} shows the results. The uncertainties were obtained applying the error propagation when computing the distance from the distance modulus.

\section{Extinction towards the RC features}

We need to exclude the possibility that the double RCs may be due to populations at different mean distances. Thus, we computed the extinction to each of them, assuming that they correspond to RC stars with the same intrinsic colours. We used the technique explained in Sect.\,4.1 of \citet{Nogueras-Lara:2018ab}. Namely, we determined the line that indicates the 50\,\% probability of membership to each of the features within the parallelograms in Fig.\,\ref{CMDs} using the GaussianMixture function \citep{Pedregosa:2011aa}. We then created a grid of extinction values ($A_{1.61}$, extinction at $1.61$\,$\mu$m) in steps of 0.01\,mag, and calculated the corresponding colour ($H-K_s$) for RC stars \citep[assuming a Kurucz model for a RC star, as explained in ][]{Nogueras-Lara:2018ab}, applying the extinction curve of \citet{Nogueras-Lara:2020aa}. We compared the colours from the stars in each of the features with the grid to obtain the best fit minimising a $\chi^2$. Table\,\ref{table4} shows the results. The mean extinction values were obtained applying a Gaussian fit to the $A_{1.61}$ distributions. The statistical uncertainties correspond to the error of the mean, and the systematics were computed varying all the parameters involved in the calculation \citep[][]{Nogueras-Lara:2018ab}. We obtained that the extinctions for both RCs are compatible within the uncertainties. This confirms that the stars are placed at roughly the same distance and the double RC is due to the SFH.

\begin{table}
\caption{Extinction towards the RC features.}
\label{table4} 
\begin{center}
\def\arraystretch{1.3}
\setlength{\tabcolsep}{3.pt}

\begin{tabular}{ccc}
\hline 
\hline 
Region & RC$_1$ & RC$_2$\tabularnewline
 & (mag) & (mag)\tabularnewline
\hline 
50 km/s & 1.85 $\pm$ 0.01 $\pm$ 0.11(0.02) & 1.81 $\pm$ 0.01 $\pm$ 0.09(0.02)\tabularnewline
20 km/s & 1.97 $\pm$ 0.01 $\pm$ 0.12(0.02) & 1.94 $\pm$ 0.01 $\pm$ 0.11(0.03)\tabularnewline
The Brick & 1.93 $\pm$ 0.02 $\pm$ 0.12(0.02) & 1.94 $\pm$ 0.03 $\pm$ 0.13(0.05)\tabularnewline
Control 1 & 1.95 $\pm$ 0.01 $\pm$ 0.12(0.02) & 1.93 $\pm$ 0.01 $\pm$ 0.12(0.02)\tabularnewline
Control 2 & 1.92 $\pm$ 0.01 $\pm$ 0.11(0.02) & 1.88 $\pm$ 0.01 $\pm$ 0.10(0.02)\tabularnewline
\hline 
\end{tabular}

\end{center}
\footnotesize
\textbf{Notes.} The subindices 1 and 2 indicate the bright and the faint RC features, respectively. The numbers in brackets correspond to the relative systematic uncertainty between values from the same region. They only consider different temperature, surface gravity, and metallicity for the RC stars. 

 \end{table}

\section{RC fitting}

We computed the average reddening free magnitude for the bright RC for each region to estimate its distance. We defined a common selection criterion for all the regions to select the RC stars ($H-K_s\in[1.3-2.0]$ and delimited by the lines  $K_s>1.2(H-K_s)+12.44$ and $K_s>1.2(H-K_s)+13.84$). In this way, we avoided any systematic effect related to the star selection. We de-reddened the stars individually applying the equation \citep[e.g.][]{Nogueras-Lara:2020aa}:

\begin{equation}
\label{der}
K_{s0} = K_s-\frac{H-K_s-(H-K_s)_0}{A_H/A_{K_s}-1} \hspace{0.5cm},
\end{equation}

\noindent where $K_{s0}$ is the reddening free magnitude; $K_s$ and $H$ are the observed magnitudes, ($H-K_s)_0$ is the intrinsic colour of RC stars \citep[$0.07\pm0.03$, e.g.][]{Nogueras-Lara:2019ad}, and $A_H/A_{K_s}$ is the extinction curve \citep[$1.84\pm0.03$,][]{Nogueras-Lara:2020aa}. 

\begin{figure}
   \includegraphics[width=\linewidth]{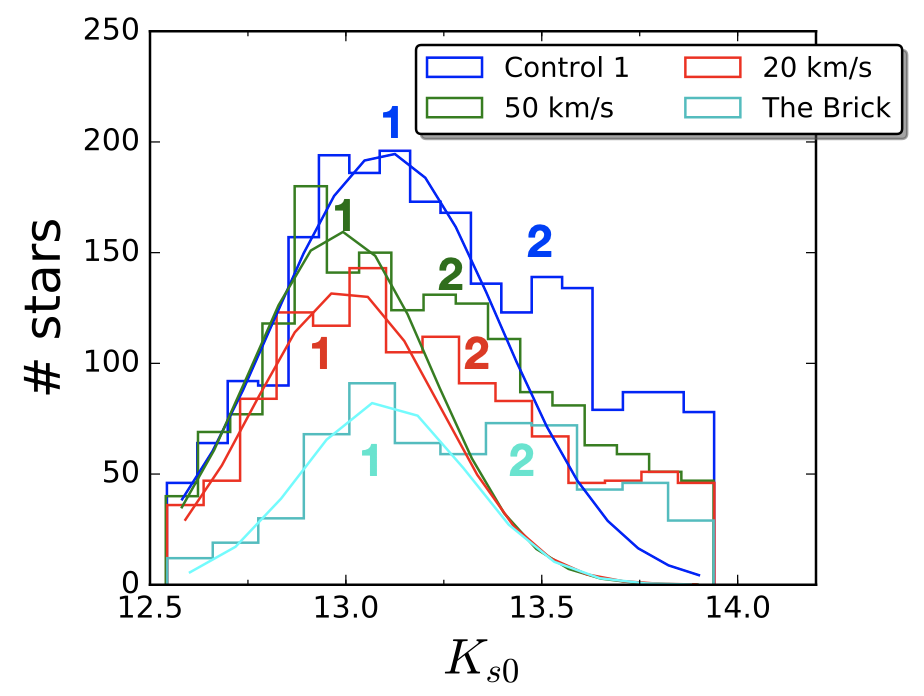}
   \caption{Histograms corresponding to the de-reddened RC stars in the analysed regions. The solid lines show a Gaussian fit to the bright RC. The numbers 1 and 2 indicate the double RC features observed in the data.}

   \label{CMD_fit}
    \end{figure}

Figure\,\ref{CMD_fit} shows the underlying distributions for the analysed regions, where we identified the double RC structure obtained in the previous section. We applied a Gaussian fit to the brightest RC peak, corresponding to the dominant old stellar population in the NSD. Table\,\ref{table3} shows the obtained results. The statistical uncertainties were negligible given the number of stars. The systematics were estimated considering all the uncertainties of the quantities involved in Eq.\,\ref{der} and equals to $\sim0.13$\,mag. To compute relative distances, the systematic uncertainties are decreased by $\sim25\,\%$, given that only the ZP systematics affect the relative difference. 

We obtained that the distance to the target clouds are compatible within the uncertainties with the distance to the control fields and also with the GC distance of $\sim8$\,kpc \citep[e.g.][]{Gravity-Collaboration:2018aa,Do:2019aa}.

\begin{table}
\caption{Average RC magnitudes and distance.}
\label{table3} 
\begin{center}
\def\arraystretch{1.3}
\setlength{\tabcolsep}{3.8pt}

\begin{tabular}{ccc}
 &  & \tabularnewline
\hline 
\hline 
Region & $K_{s\_RC}$ & distance\tabularnewline
 & (mag) & (kpc)\tabularnewline
\hline 
50 km/s & 12.99 & 8.0 $\pm$ 0.5\tabularnewline
20 km/s & 13.00 & 8.0 $\pm$ 0.5\tabularnewline
The Brick & 13.10 & 8.4 $\pm$ 0.5\tabularnewline
Control 1 & 13.10 & 8.4 $\pm$ 0.5\tabularnewline
Control 2 & 13.07 & 8.3 $\pm$ 0.5\tabularnewline
\hline 
\end{tabular}
\end{center}
\footnotesize
\textbf{Notes.} $K_{s\_RC}$ indicates average value from the Gaussian fits. The associated systematic uncertainty is $0.13$\,mag in all the cases. For relative comparison the uncertainty is $0.1$\,mag. 

 \end{table}

\section{Discussion and conclusion}

In this letter we computed the distances to three molecular clouds in the CMZ (50 km/s, 20 km/s, and the Brick clouds) and investigated their CMZ membership. 

We obtained that all fields towards the clouds present a double RC with equal magnitude difference ($\Delta K_s$) between both features, indicating similar stellar populations.  The $\Delta K_s$ and the slope of the features agree with previous work for the NSD ($\Delta K_s = 0.44\pm0.01$\,mag from \citet{Nogueras-Lara:2019ad}, and a slope of $1.19\pm0.04$, obtained transforming the extinction ratio $A_H/A_{K_s}=1.84\pm0.03$ from \citealt{Nogueras-Lara:2020aa}). Therefore, the stars present in the target molecular clouds are compatible with the extinction curve and the star formation history found for the NSD, and so the control regions. If the clouds were located closer to Earth, outside of the NSD, then they should show single RCs, as observed for the inner bulge control field. This indicates that none of the target clouds likely belong to the inner bulge. Moreover, we also computed the distance to the clouds applying a linear fit to the bright RCs (Eq.\ref{equation_cut}), and found that they are compatible with the GC distance of $\sim8$\,kpc.

We investigated the nature of the two RC features computing the extinction towards each of them. We obtained that the extinction towards both features is compatible within the uncertainties in all the cases. Therefore, it is very unlikely that they are placed at different distances. However, the Brick was recently claimed to not belong to the CMZ, having a distance from the Sun of 7.20$\pm$0.16$\pm$0.2\,kpc \citep{Zoccali:2021aa}. If this were the case, the stellar population belonging to the NSD according to \citet{Zoccali:2021aa} (in this letter the secondary RC) would have a significantly larger extinction given the high amount of gas and dust present in the CMZ. Thus, obtaining similar extinctions for both RC features, and the very similar values in comparison to other regions belonging to the NSD (control regions), indicate that the Brick is likely at the GC distance. Furthermore, the Brick appears to be less transparent in the near infrared than the other analysed clouds (see Fig.\ref{CMDs}), being very unlikely to observe stars beyond it. Hence, observing the features of the star formation expected for the NSD, places the Brick at the GC.

\begin{figure}
   \includegraphics[width=\linewidth]{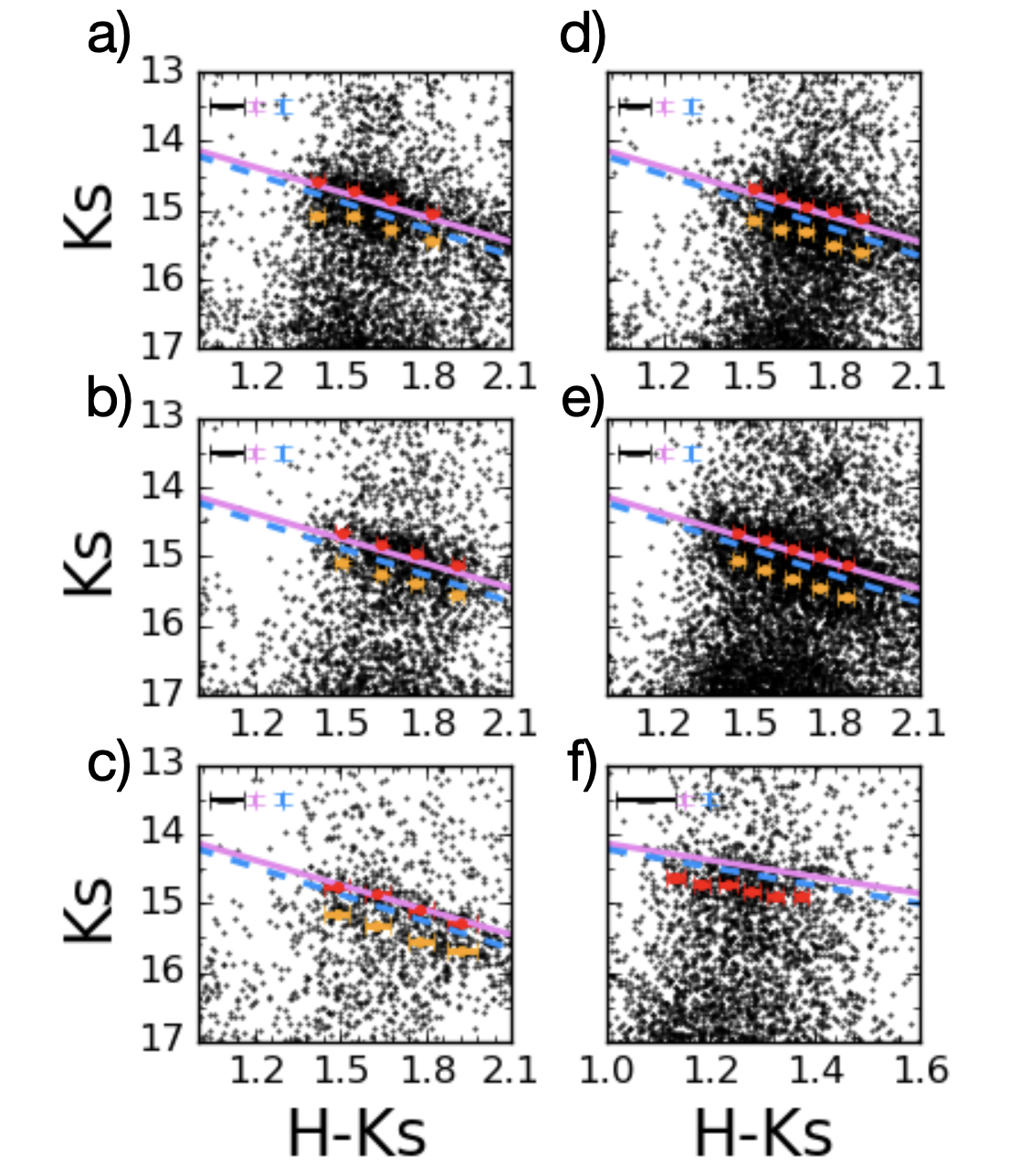}
   \caption{Colour magnitude diagrams $K_s$ vs. $H-K_s$ of the target regions: a) 50 km/s cloud. b) 20 km/s cloud. c) The Brick cloud. d) Control 1. e) Control 2. f) Bulge region. The red and orange points indicate the detected double feature for the RC. The violet solid line corresponds to the reddening line of RC stars at 8.2\,kpc distance following the extinction curve by \citep{Nogueras-Lara:2020aa}. The blue dashed line shows the extinction curve by \citet{Minniti:2020aa}. The black, violet, and blue error-bars correspond to the systematics of the data, and the used extinction curves, respectively.}

   \label{CMD_lines}
    \end{figure}

To further analyse whether the Brick and the other clouds belong to the CMZ, we used the same approach as \citet{Zoccali:2021aa}. We assumed reddening free ($H-K_s$, $K_s$) values of a RC as inferred from the results obtained for the Baade's window (0.22,13.2), and plotted the extinction curve from that starting point \citep{Zoccali:2021aa}. If the RC stars in our CMDs lie at the GC, then they should scatter around this reddening line. We assumed the extinction curve used in \citet{Zoccali:2021aa}, $A_{K_s}= 1.308\pm0.05\times E(H-K_s)$ \citep{Minniti:2020aa}, and the extinction curve $A_{K_s}= 1.19 \pm0.04 \times E(H-K_s)$  \citep{Nogueras-Lara:2020aa}. Figure\,\ref{CMD_lines} shows the results. For all the regions, but the inner bulge one, we found that the bright RC is fitted well by the two reddening lines. The line obtained using the extinction curve by \citet{Minniti:2020aa} is below the other one, but yet compatible within the uncertainties. Only the inner bulge RC is below the lines, indicating a probably different average distance (larger according to our results). Thus, we concluded that all the regions in the NSD actually belong to the NSD and are within the CMZ, including the Brick cloud.

We believe that the main causes for the disagreement between our result and the one of \citet{Zoccali:2021aa} are: {\bf 1)} It is necessary to consider the double red clump features given the star formation history in the NSD to avoid the miss-identification of any of the clumps. {\bf 2)} The extinction curve by \citet{Minniti:2020aa} was derived using classical cepheids in the bulge and the far side of the disc, but not in the GC. Hence, the line-of-sight towards the GC is not adequately covered, so possible variations of the extinction curve along the line-of-sight would make this law not appropriate for the GC region. On the other hand, previous work on the NSD \citep{Nogueras-Lara:2019ac,Nogueras-Lara:2020aa}, obtained an extinction curve compatible with the slopes of the analysed RC features, so it might be more appropriate to study the CMZ. Our results agree for both extinction curves, but an underestimation of the uncertainties or not considering the double RC features, might produce biased results. Calculating the expected position for the RC at the GC distance using both extinction curves gives a difference of $\sim0.18$\,mag for $H-K_s=1.75$\,mag. Therefore, the $K_s$ distance of $\sim0.28$\,mag measured by \citet{Zoccali:2021aa} between the bright RC and the theoretical reddening line, would be reduced to $\sim0.1$\,mag when using the extinction curve by \citet{Nogueras-Lara:2020aa} (being then in agreement with the GC distance within the uncertainties). {\bf 3)} An underestimation of the uncertainties. The slope of the extinction curve used is $1.308\pm0.05$. Therefore, to compute the $K_s$ magnitude corresponding to a given $H-K_s$ value, the equation is $K_s = 1.308\times (H-K_s) + 12.91$, where the intercept was computed using the Baade's window starting point. Thus, the error for a computed $K_s$ value (assuming no uncertainty for the intercept) would be $\Delta K_s = \sqrt{ (\partial K_s/\partial slope )^2\Delta slope^2  }$ $=\sqrt{ (1.308\times0.05 )^2  } \sim0.07$\,mag. Therefore, after error propagation, that implies a systematic uncertainty of $\sim300$\,pc instead of $200$\,pc \citep{Zoccali:2021aa}. Also, we think that \citet{Zoccali:2021aa} only considered the systematics of the $K_s$ ZP, but not of the $H$ ZP in their work. Taking both sources of uncertainty into account would increase the uncertainty up to $\sim0.1$\,mag, corresponding to $\sim350$\,pc at a distance of 7.2\,kpc ($\sim400$ at 8\,kpc).
 
Finally, we computed the distance towards all the regions in the NSD de-reddening RC stars selected in a similar way for all the regions to avoid selection effects and obtain that the distance to all the regions is compatible between them with an uncertainty of $\sim400$\,pc (Table\,\ref{table3}). Therefore, we concluded that the studied clouds are compatible with the GC distance.

  \begin{acknowledgements}
      This work is based on observations made with ESO
      Telescopes at the La Silla Paranal Observatory under programme
      IDs 195.B-0283. We thank the staff of
      ESO for their great efforts and helpfulness. We thank Steven Longmore for useful discussion and comments on the manuscript. F.N.-L. and N.N. gratefully acknowledge support by the Deutsche Forschungsgemeinschaft (DFG, German Research Foundation) – Project-ID 138713538 – SFB 881 ("The Milky Way System", subproject B8). R.S. acknowledges financial support from the State
Agency for Research of the Spanish MCIU through the "Center of Excellence Severo
Ochoa" award for the Instituto de Astrof\'isica de Andaluc\'ia (SEV-2017-0709). R.S.  acknowledges financial support from national project
PGC2018-095049-B-C21 (MCIU/AEI/FEDER, UE).
\end{acknowledgements}

\bibliography{../BibGC.bib}
\end{document}